# Observation of a Non-Hermitian Phase Transition in an Optical Quantum Gas

Fahri Emre Öztürk[1], Tim Lappe[2], Göran Hellmann[1], Julian Schmitt[1,*], Jan Klaers[1,**], Frank Vewinger[1], Johann Kroha[2], and Martin Weitz[1,*]

[1]*Institut für Angewandte Physik, Universität Bonn, Wegelerstr. 8, 53115 Bonn, Germany*

[2]*Physikalisches Institut, Universität Bonn, Nussallee 12, 53115 Bonn, Germany*

[*]Corresponding authors. E-mail: schmitt@iap.uni-bonn.de (J. S.);
martin.weitz@uni-bonn.de (M. W.)

**Abstract:** Quantum gases of light, as photons or polariton condensates in optical microcavities, are collective quantum systems enabling a tailoring of dissipation from e.g. cavity loss. This makes them a tool to study dissipative phases, an emerging subject in quantum manybody physics. Here we experimentally demonstrate a non-Hermitian phase transition of a photon Bose-Einstein condensate to a new dissipative phase, characterized by a biexponential decay of the condensate's second-order coherence. The phase transition occurs due to the emergence of an exceptional point in the quantum gas. While Bose-Einstein condensation is usually connected to ordinary lasing by a smooth crossover, the observed phase transition separates the novel, biexponential phase from both lasing and an intermediate, oscillatory condensate regime. Our findings pave the way for studies of a wide class of dissipative quantum phases, for instance in topological or lattice systems.

Creating and understanding phases of systems which are dissipatively coupled to the environment is of importance in research fields ranging from optics to biophysics (1-6). One intriguing aspect of this openness is the possible existence of quantum states which would not be accessible otherwise (7-10). Near-equilibrium physics (11,12) has been studied in optical quantum gases (13), such as photons or polaritons - strongly coupled, mixed states of light and matter - despite their driven-dissipative nature. In particular, Bose-Einstein condensates of photons, realized in dye-filled microcavities by multiple photon absorption and re-emission cycles, provide a platform to study quantum dynamics in an open, grand canonical situation where the condensate particles are coupled to a reservoir of the photo-excitable dye molecules (14). Photon condensates have the macroscopic mode occupation in common with lasers, but operate near thermal equilibrium, in distinct contrast to lasers. Naively, see also Refs. (8,15), one would expect a smooth crossover between lasing and condensation given that both phenomena exhibit spontaneous symmetry breaking.

Recently, two reports observing oscillatory dynamics in open dye microcavity systems have appeared (6, 16), a phenomenon that at large resonator losses crosses over to the relaxation oscillations known in laser physics. Other than in a laser, the stochastic driving induced by grand canonical condensate fluctuations make the system dynamics observable in stationary-state operation, which characterize the system's state by its second-order coherence. In contrast to closed systems governed by time-reversal symmetric, i.e., Hermitian dynamics, the dissipative coupling to the environment is described by a non-Hermitian time-evolution operator with complex eigenvalues. Of special interest are exceptional points, where the eigenvalues and the corresponding eigenmodes coalesce (1,17-20). Such points are well known to enable phase transitions (21); see refs. (8,22) for a recent proposal of a first-order phase transition between a photon laser and a polariton condensate.

Here we report the observation of a non-Hermitian phase transition to a novel dissipative phase in a photon Bose-Einstein condensate coupled to the environment. It occurs at an exceptional point which resides well inside the condensed regime, at loss rates a thousand times smaller than the crossover to lasing. The revealed phase of the optical quantum gas, with biexponential decay of the second order coherence, is thus separated from both the oscillatory condensate phase and lasing by a true phase transition.

To prepare an open photon Bose-Einstein condensate coupled to a reservoir, we use a dye microcavity apparatus (11,23-25), see Fig.1A for a schematic. The short mirror spacing of a few wavelengths discretizes the longitudinal wavevector, such that only modes with a fixed longitudinal mode number are accessible to the photon gas at room temperature. This imposes a quadratic dispersion as function of the transverse wave numbers, and the photon gas becomes formally equivalent to a harmonically trapped two-dimensional gas of massive bosons, which supports Bose-Einstein condensation (25). Photons are injected by pumping with a laser beam. They thermalize to the dye temperature by absorption-re-emission cycles, before being lost by e.g. mirror transmission (Fig.1A, right panel). The used rhodamine dye fulfils the Kennard-Stepanov relation $B_{em}/B_{abs} \propto e^{-\hbar\omega/k_B T}$, a Boltzmann-type frequency scaling between the Einstein coefficients for absorption $B_{abs}$ and emission $B_{em}$. Experimental spectra are shown in Fig.1B, showing agreement with an equilibrium Bose-Einstein distribution within experimental accuracy.

The steady-state particle flux from the pump beam through the dye microcavity condensate and out to the environment induces a novel behavior of the particle number fluctuations, which we analyze next. In this open system, the sum $X$ of the condensate photon number $n(t)$ and dye molecular excitations $M_e(t)$ is conserved only on average (14,25), $\bar{X} = \bar{n} + \bar{M}_e = const,$ where the bar denotes the time average. The dynamics of the corresponding fluctuations $\Delta n$ and $\Delta X$ around the mean can be derived from a Lindblad equation that incorporates the thermally driven fluctuations of the grand canonical system (equivalent to a Langevin equation). For small deviations $\Delta$n and $\Delta$X, this leads to a set of equations (25):

$$\frac{d}{dt}\begin{pmatrix} \Delta n \\ \Delta X \end{pmatrix} = \mathbb{A} \begin{pmatrix} \Delta n \\ \Delta X \end{pmatrix}, \tag{1}$$

with the non-Hermitian matrix

$$\mathbb{A} = \begin{pmatrix} -2\delta & \omega_0^2/\kappa \\ -\kappa & 0 \end{pmatrix}, \tag{2}$$

where $\delta = \frac{1}{2}B_{em}(\tilde{M}_e/\bar{n} + \bar{n})$ is the damping rate of the photon number fluctuations, $\omega_0 = \sqrt{\kappa B_{em}\bar{n}}$ an oscillation frequency, and the loss rate $\kappa$ models photon loss. It is instructive to first discuss the expected response to an instantaneous fluctuation at a time $t_0$. With the exponential ansatz $(\Delta n_0, \Delta X_0) \cdot e^{\lambda(t-t_0)}$ one obtains solutions characterized by the matrix eigenvalues $\lambda_{1,2} = -\delta \pm \sqrt{\delta^2 - \omega_0^2}$. For a damping $\delta$ below the natural angular frequency $\omega_0$ of the undamped system, the eigenvalues become complex, corresponding to a (damped) oscillatory solution, while in the opposite regime of a large damping ($\delta > \omega_0$) we arrive at real eigenvalues implying a biexponential decay. At $\delta = \omega_0$ the eigenvalues and the corresponding solutions coalesce, marking an exceptional point. For stationary conditions, i.e. constant pumping and loss, the dynamics of the grand canonical system driven by thermal fluctuations becomes stochastic; the modes described above can thus only be observed in the number correlations of the condensate mode, described by $g^{(2)}(\tau) = 1 + e^{-\delta\tau}\left(C_1 e^{-\sqrt{\delta^2-\omega_0^2}\tau} + C_2 e^{\sqrt{\delta^2-\omega_0^2}\tau} + c.c.\right)$, with constants $C_1$ and $C_2$ (25).

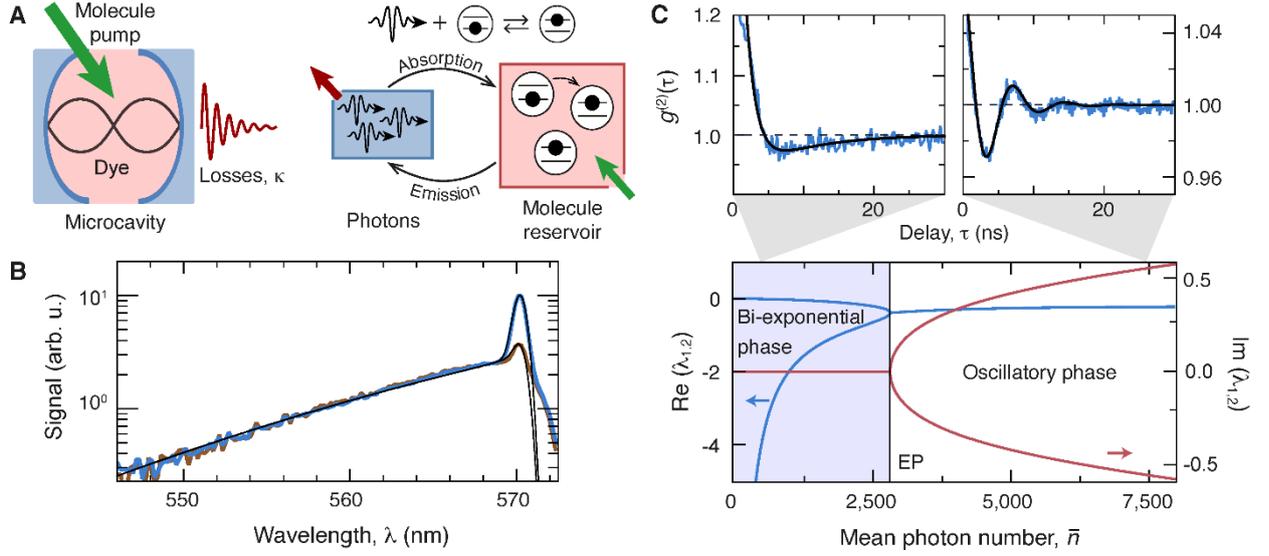

**Fig. 1: Experimental principle.** **(A)** Photons are trapped within a dye-filled microcavity, where losses $\kappa$ are compensated by pumping the dye molecules with a laser. The photon gas is coupled to this reservoir by the exchange of excitations between photons and electronically excited dye molecules (right panel). **(B)** Spectra of the emission for average photon numbers $\bar{n} \cong 2100$ and $10300$, showing a thermalized photon gas with a condensate peak at the position of the low-energy cutoff, closely following the expected (experimentally broadened) Bose-Einstein distributions at 300K (black lines). **(C)** Second-order correlations $g^{(2)}(\tau)$ of the condensate, recorded at $\bar{n} \cong 2300$ (left) and $\cong 14000$ (right) respectively, with fitted theory curves (25) (black lines), showing oscillatory behavior for large and a biexponential decay for small photon numbers. The bottom panel shows predictions of real (blue) and imaginary (red) parts of the eigenvalues $\lambda_{1,2}$ (for a molecule number $M=5 \cdot 10^9$), which are real below and complex above the exceptional point (EP).

Tuning between the different regimes - damped oscillatory for $\delta < \omega_0$ and biexponentially decaying correlations for the opposite case - experimentally is achieved by varying the average photon number $\bar{n}$, which with $\omega_0 = \omega_0(\bar{n})$, $\delta = \delta(\bar{n})$ serves as a control parameter. For small variations of $\bar{n}$, the photon condensate will remain in the oscillatory or the bi-exponential regime when being far from the exceptional point ($\delta \ll \omega_0$ or $\delta \gg \omega_0$). At the exceptional point ($\delta = \omega_0$), however, the condensate dynamics, as observed in $g^{(2)}(\tau)$, becomes very sensitive to changes in $\bar{n}$, and may change qualitatively abruptly. We attribute the exceptional point as marking a non-Hermitian phase transition, separating two dynamical condensate phases. The phase transition mechanism draws analogies with that of quantum phase transitions in closed systems, where typically two energy eigenvalues cross; see also Ref.26 for a proposal of a dissipative phase transition more closely resembling that of (usual) Hermitian systems. For our system, deep in one of the two condensate phases the eigenvalues of the fluctuation matrix $A(\bar{n})$ are gapped in the complex plane and we have $Re(\lambda_1-\lambda_2)\neq0$ or $Im(\lambda_1-\lambda_2)\neq0$ on different sides the transition respectively (see Fig.1C, bottom panel). The gap closes ($\lambda_1=\lambda_2$) at the exceptional point. Interestingly, critical fluctuations known from equilibrium phase transitions are here replaced by an enhanced sensitivity in the correlation dynamics to changes of the control parameter at the phase boundary. Note that no spontaneous symmetry breaking occurs, which is a property shared with e.g. the fermionic Mott-Hubbard transition. Thermal, reservoir-induced fluctuations of the photon condensate are crucial for the emergence of the described non-Hermitian phase transition.

To experimentally determine the second-order coherence of the photon condensate around the exceptional point, the microcavity emission passes a mode filter to separate the condensate mode from the higher transverse modes. The transmitted light is polarized and directed onto a fast photomultiplier, whose electronic output allows for correlation analysis. Typical obtained traces of the second order-order correlations are shown in Fig.1C for a cutoff wavelength $\lambda_c \cong 571.3$ nm. While for the larger condensate photon number of $\bar{n} \cong 14000$ the second-order coherence is oscillatory (right panel), for the smaller photon number of $\bar{n} \cong 2100$ it exhibits biexponential behavior (left panel), in good agreement with theory. The difference of damping constant and undamped oscillation frequency, as determined from the fits, is $(\delta-\omega_0)/2\pi = -99(7)$ MHz and 19(2) MHz for the two datasets, and consequently the data can be assumed to be in the oscillatory phase for the former and in the biexponential condensate phase for the latter dataset. The presence of both the thermal cloud and the condensate peak in the observed spectra (Fig.1B), which are a consequence of Bose-Einstein (quantum) statistics, are attributed as evidence for the quantum manybody character of the phases.

Figure 2A schematically shows the hierarchy of phases for fixed values of the average photon and molecule number, and Fig.2B gives a three-dimensional plot of the full expected phase diagram. The indicated crossover between lasing and condensation occurs when for $\kappa >> \overline{M}_g B_{abs}$ the loss rate becomes so large that photons leak from the cavity before thermalizing via reabsorption (27,28). The phase transition between the intermediate oscillatory and the biexponential phases for $\delta = \omega_0$ occurs at $\sqrt{\kappa B_{em} \overline{n}} \simeq \frac{1}{2} B_{em}(\overline{M}_e/\overline{n} + \overline{n})$, and features a grand canonical $(\overline{M}_e >> \overline{n}^2)$ and a canonical branch $(\overline{M}_e << \overline{n}^2)$ of the phase boundary, corresponding to the first or second term in the sum being dominant. Here, the first (grand canonical) term, understood to arise from retrapping of spontaneous emission, is absent in usual laser theory. Experimentally, with $\kappa/(\overline{M}_g B_{abs}) \cong 1.1 \cdot 10^{-3}$ photon thermalization dominates over photon loss, i.e. the exceptional point is clearly in the Bose-Einstein condensed regime.

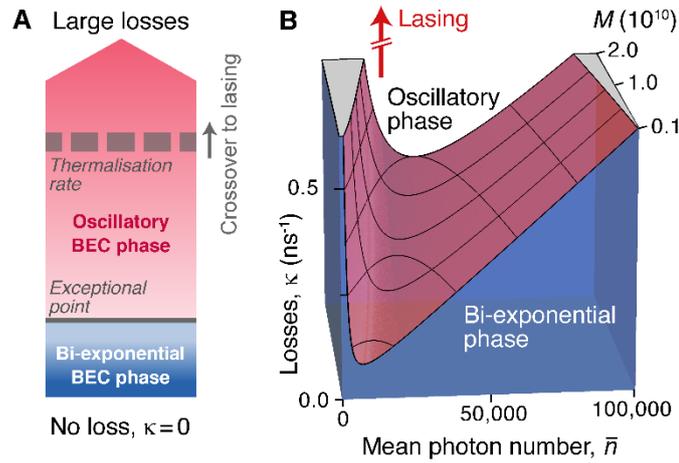

**Fig. 2: Expected phase diagram. (A)** Illustration of the hierarchy of phases for increasing losses, with fixed condensate size and molecule number. The exceptional point introduces a well-defined phase boundary between photon Bose-Einstein condensates with weakly-dissipative (biexponential) and dissipative (oscillatory) correlation dynamics. At loss rates exceeding the thermalization rate, a crossover connects the oscillatory phase to the lasing regime. **(B)** Calculated phase boundary ($\delta = \omega_0$) between the two condensate phases (25), as a function of mean condensate and molecule numbers.

To explore the phase transition between the two different condensate phases, we have recorded the photon number correlations at different average photon numbers. Figure 3A shows the variation of $\delta$-$\omega_0$, as determined from the fits of the correlation data. While for condensate sizes above $\overline{n}_{EP} \cong 2800$ the second-order coherence shows a damped oscillation, for smaller photon numbers the data exhibits a biexponential decay of the correlations. Figures 3B and 3C show the obtained decay times and, for the case of the oscillatory phase, the oscillation frequency depending on the condensate size. Both data sets give evidence for the photon condensate to undergo a non-Hermitian phase transition to the biexponential phase at a critical condensate occupation $\bar{n}_{EP}$. The deviation of the observed decay times from the prediction for short times is attributed to the 500ps resolution of the detection system. Notably, when approaching the phase transition from below, the two characteristic decay times merge towards a single one, and when approaching the transition from above, the oscillation frequency converges to zero. This is in good agreement with the expectation that at the exceptional point, a (single) exponential decay of the second-order coherence occurs due to the coalescence of the two eigenvalues, $\lambda_1=\lambda_2=\delta$. The revealed phase transition is visible in the temporal correlations, but not in the average values.

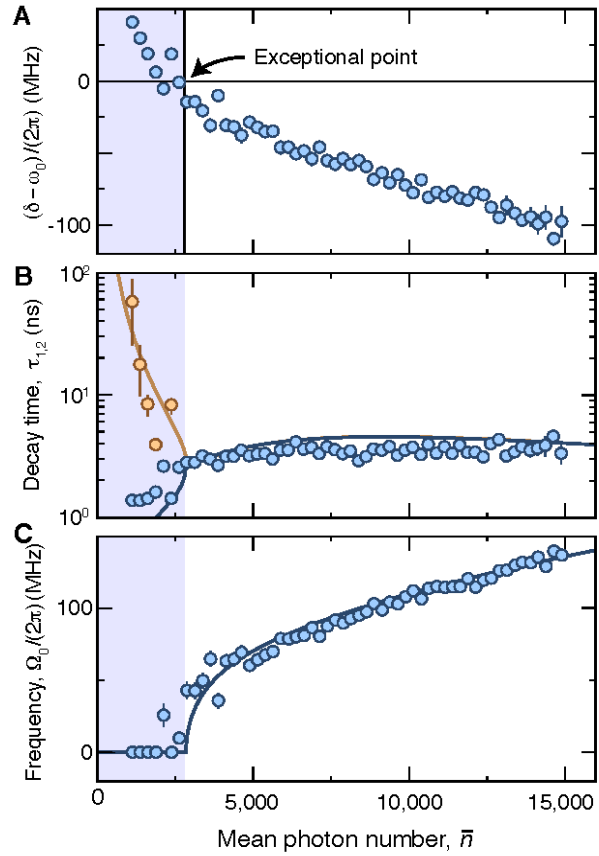

**Fig. 3: Non-Hermitian phase transition. (A)** Variation of the difference δ-ω₀, **(B)** the decay times, and **(C)** the oscillation frequency with average photon number, determined from the correlation data. Negative (positive) values of δ-ω₀ **(A)** indicate an oscillatory (biexponential, blue shading) coherence function, respectively. With increasing condensate size, the two decay times **(B)** merge towards a single value near a photon number of $\bar{n}_{EP} \cong 2800$, which displays the phase transition expected from δ-ω₀=0 in **(A)**. Accordingly, above $\bar{n}_{EP}$ the oscillation frequency becomes non-vanishing. The fits yield κ≅2.2(2) ns⁻¹, M≅4.76(3)·10⁹. Error bars are calculated from the uncertainties of the fit parameters. (Cutoff wavelength: λ𝒸= 571.3nm).

Next, we have recorded data at different cavity low-frequency cutoffs and dye concentrations, to explore the phase diagram to beyond a single control parameter. The resulting change of the wavelength of condensate photons modifies both the loss as well as the Einstein coefficient. Due to the shape of the phase boundary at $\delta=\omega_0$, upon rescaling the photon number as $\bar{n}/\sqrt{\bar{M}_e}$ and the loss rate as $\kappa/B_{em}\sqrt{\bar{M}_e}$, the phase diagram in Fig. 2B collapses to a two-dimensional one (25). Corresponding data is summarized in Fig.4A. To obtain $\kappa$ and the molecule number $M$ curves similar as for a single data set shown in Fig.3 were fitted to all data. Our experimental data maps out the non-Hermitian phase transition between the oscillatory and biexponential phase within the investigated parameter range, in good agreement with expectations (black line). The variation of the normalized decay times and oscillation frequency versus the scaled loss rate in Fig.4B for a fixed value of $\bar{n}/\sqrt{\bar{M}_e} \simeq 0.27$ demonstrate the branching of the eigenvalues when reducing the loss towards the idealized case of a perfect photon box.

To conclude, the state of a macroscopic quantum system on different sides of an exceptional point can be in two distinct regimes. We have observed the associated dissipative phase transition from an oscillatory to a biexponential dynamical phase of a dye microcavity photon Bose-Einstein condensate, and mapped out the corresponding phase diagram. This reveals a state of the light field, which, contrary to the usual picture of Bose-Einstein condensation, is separated by a phase transition from the phenomenon of lasing.

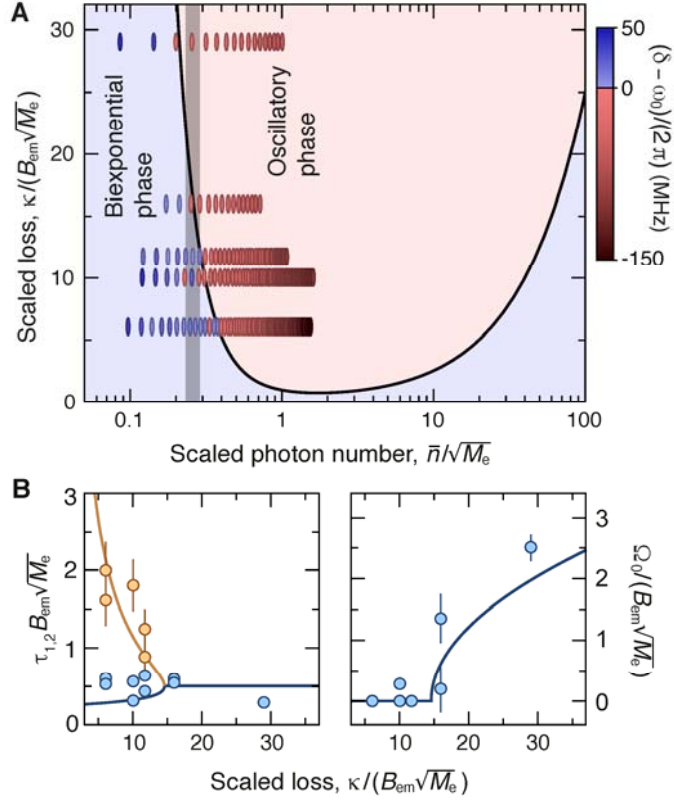

**Fig.4: Exploring the phase diagram. (A)** The phase boundary between the two regimes (solid line) is mapped out by recording different data sets with different cutoff wavelengths and dye concentrations. From the corresponding coherence functions (as in Fig.3) we identify a condensate in the biexponential (blue points) or oscillatory (red points) phase, respectively. **(B)** Variation of the normalized decay times (left) and oscillation frequency (right) with loss for a scaled photon number $\bar{n}/\sqrt{M_e} = 0.27(2)$ (grey shaded area in **(A)**), showing the transition into the biexponential phase when reducing losses, in good agreement with theory (solid lines). Error bars are calculated from the uncertainties of the fit parameters.

For the future, the demonstrated mechanism for the realization of new states of the light field is expected to lead to a wide class of dissipative phases in lattice systems (8,29). In two-dimensional coupled cavity arrays, also new dissipative topological phases can emerge (30). A further intriguing perspective is to search for critical fluctuations in the non-Hermitian system in the vicinity of an exceptional point (22).

**Acknowledgments:** We thank S. Diehl, M. Scully, and H. Stoof for discussions. **Funding:** We acknowledge support by the DFG within SFB/TR 185 (277625399) and the Cluster of Excellence ML4Q (EXC 2004/1–390534769), the EU within the Quantum Flagship project PhoQuS, and the DLR with funds provided by the BMWi (50WM1859). J.S. thanks the University of Cambridge for support during the early stages of this work, and M.W. the CAIQuE for support of a guest stay at UC Berkeley. **Data and materials availability:** Data shown in the figures are available on the Zenodo database (31).

*\*\*Present address: Complex Photonic Systems (COPS), MESA+ Institute for Nanotechnology, University of Twente, 7522 NB Enschede, Netherlands.*

## Supplementary Materials:

### Materials and Methods

<u>Experimental system</u>

The used experimental apparatus is similar to as described in earlier works, see Refs. 6,14 of the main text for details. Our optical microcavity consists of two ultra-high reflectivity mirrors with reflectivity above 99.998% of R=1m curvature spaced by a distance of $D_0 \cong 1.4 \mu m$. The cavity is filled with rhodamine dye dissolved in ethylene glycol (concentration: $10^{-3}$ mol/l). At room temperature, rapid transverse decoherence caused by collisions of solvent molecules with the dye suppresses the formation of polaritons (32,33). Importantly, this also prevents Rabi oscillations of the molecules, meaning that the oscillations of the photon number observed in the oscillatory condensate regime cannot be attributed to a coherent evolution, but are rather due to the weakly dissipative character of the dye microcavity system. The small mirror spacing causes a large frequency spacing between adjacent longitudinal optical modes that is comparable with the emission width of the dye molecules. In this regime we observe that to good accuracy the resonator is only populated with photons of a fixed longitudinal mode, q=7 here, and the two transverse modal degrees of freedom make the system two-dimensional. The dye microcavity is pumped with a laser beam of 532 nm wavelength at an angle near 45° with respect to the optical axis. To suppress pumping of the dye into triplet states and excessive heating, the pump pulses are acousto-optically chopped into 500ns long pulses at a 50 Hz repetition rate.

In the cavity the dispersion relation due to the short mirror spacing is modified with respect to free space and acquires quadratic, i.e. massive particle-like, character. The transverse $TEM_{00}$ mode becomes the lowest populated eigenmode, which acts as a low-frequency cutoff at energy $\hbar\omega_c = hc/\lambda_c$, where $\lambda_c$ is the cutoff wavelength. Further, the mirror curvature leads to harmonic confinement of the photon gas, making the photons equivalent to a two-dimensional, harmonically confined gas of massive bosons with effective mass $m_{eff} = \hbar\omega_c/(c/n)^2$, where c denotes the speed of light and $n \cong 1.43$ is the refractive index of the solvent ethylene glycol. For such a system it is known that a BEC exists at thermal equilibrium conditions (34).

Thermal equilibrium of photons in the cavity is achieved as the photons are absorbed and re-emitted many times by the dye molecules. The used rhodamine dye fulfills the Kennard-Stepanov relation to good accuracy (35). This universal thermodynamic frequency scaling between absorption and emission is well known to apply for systems with rovibrational spectra on top of both lower and upper electronic levels in equilibrium. The conversion of photons into dye electronic excitations and vice versa (right panel of Fig. 1A of the main text) can be seen as an exchange of both energy and particles with the dye which acts as a reservoir in the grand canonical sense. These multiple absorption and re-emission processes induce a thermal spectral distribution of the photon gas at the temperature of the dye rovibrational excitations, which are at room temperature. Given that thermal emission is negligible in the limit of $\hbar\omega_c (\cong 2.1\text{eV}) >> k_B T (\cong 1/40\text{eV})$, temperature and chemical potentials are independently tunable. This is a striking difference to the usual case of black-body radiation where photons vanish in the system walls upon lowering the temperature instead of exhibiting condensation. In the dye microcavity system, both photon gas thermalization with a freely adjustable chemical potential and Bose-Einstein condensation has been observed in earlier works, see Refs.11,23,24 for details.

To experimentally monitor the number statistics of the photon condensate, the transmission of one of the microcavity mirrors following a mode filter is directed onto a fast photomultiplier. The mode filter separates the (TEM$_{00}$) condensate mode from the higher transverse modes forming the thermal cloud by transmission through two optical pinholes acting as a real space and a momentum filter, respectively. In addition, a polarizer is placed into the detection path, removing the polarization degeneracy. The electronic signal of the used photomultiplier (Photek PMT 210) is analyzed with a fast oscilloscope, yielding a correlation signal from the time-resolved photomultiplier signal traces. This signal is used as a measure for the time-dependent second-order coherence $g^{(2)}(\tau)$ of the photon condensate. Experimental data for $g^{(2)}(\tau)$ was recorded in 5 different measurement runs, sampling dye microcavity emission signals acquired within $3 \cdot 10^4$ pump beam pulses each. In each of the measurement runs data for different pump beam powers was recorded so as to vary the condensate size, with calibration of the photon number obtained by simultaneously recording spectra of the dye microcavity emission. The spectra relate the condensate mode population to the photon number in the thermal cloud, the latter equaling the critical number $N_c = \frac{\pi^2}{3}\left(\frac{k_B T}{\hbar \Omega}\right)^2 \cong 80700$ for the used trap frequency $\Omega \cong 2\pi \cdot 40$ GHz. The data used for further analysis are averages of the second-order coherence functions for different mean photon numbers $\overline{n}$ with a number bin size of 250 photons, as shown in the top panels of Fig. 1C for two different values of $\overline{n}$. In preparatory measurements with a pulsed laser source, the time resolution of the detection system for measurements of the second-order coherence was determined to be $\cong 500$ps.

Theoretical model

The dynamics of photons coupled to the dye reservoir can be modeled with the following set of rate equations:

$$\frac{d\langle n \rangle}{dt} = \left[B_{em}\langle M_e \rangle(\langle n \rangle + 1) - B_{abs}\langle M_g \rangle\langle n \rangle\right] - \kappa\langle n \rangle, \qquad (S1a)$$

$$\frac{d\langle M_e \rangle}{dt} = -\left[B_{em}\langle M_e \rangle(\langle n \rangle + 1) - B_{abs}\langle M_g \rangle\langle n \rangle\right] + R_p\langle M_g \rangle, \qquad (S1b)$$

where $\langle n \rangle$ denotes the time-dependent expectation value of the number of photons in the condensate mode and $\langle M_g \rangle$ and $\langle M_e \rangle$ describe the time-dependent number of dye molecules in electronic ground and excited states, respectively, with $M = \langle M_e \rangle + \langle M_g \rangle$ as the total molecule number. Further, $B_{abs}$ and $B_{em}$ denote the Einstein coefficients for absorption and emission, respectively, which are related by the Kennard-Stepanov relation, $B_{em}/B_{abs}=exp(-\hbar(\omega-\omega_{ZPL})/k_B T)$, where $\omega_{ZPL} \cong 2\pi c/545$nm and $T \cong 300$K denote the frequency of the zero-phonon line for rhodamine and the temperature of the dye bath, respectively. Further, $\kappa$ denotes the photon loss and $R_p$ the pump rate, respectively. In the absence of pumping and loss, with the sum $X = \langle n \rangle + \langle M_e \rangle$ of the photon number and dye molecular electronic excitations being strictly conserved, the dynamics can be described by eq. S1a alone, predicting the fluctuation properties of grand canonical Bose-Einstein condensation, as described in previous work (36-40), see also Ref. 14. For a large relative size of the dye reservoir, that is $M_{eff} >> \bar{n}^2$, where $M_{eff} = M/(2 + 2\cosh((\omega_c - \omega_{ZPL})/k_B T))$ ($\cong \bar{M}_e$ for the detuning values chosen here) denotes the effective reservoir size, grand canonical conditions are fulfilled, and in the condensed state, photon number fluctuations of order of the average particle number emerge. In this limit, the normalized second-order photon coherence function of the condensate reaches $g^{(2)}(0) = 2$, i.e., is the same as for a thermal source. On the other hand, when the effective reservoir size is small, for $M_{eff} << \bar{n}^2$, the particle-number fluctuations reduce to a Poissonian level, i.e., they resemble the number statistics of a "usual" canonical ensemble condensate, or of a laser. By varying effective relative size of the reservoir, the second-order coherence of the photon condensate can be tuned between $g^{(2)}(0) = 1$ and $g^{(2)}(0) = 2$.

The dynamics of fluctuations of the photon number $\Delta n = \langle n \rangle - \bar{n}$ and the total number of excitations $\Delta X = \langle X \rangle - \bar{X}$ of the open system is described by eqs. 1-2 of the main text. These equations are obtained by linearizing around the average values. Remember here that $\bar{n}$ and $\bar{X}$ denote the average values (i.e. averaged over times longer than the correlation time) of the condensate mode photon number and sum of excitations respectively, while $\langle n \rangle$ and $\langle X \rangle$ denote the instantaneous expectation values at time $t$.

The response to an instantaneous fluctuation of the photon number occurring at a time $t_0$ around the average value $\bar{n}$ can be written in the form $\Delta n(t) = e^{-\delta(t-t_0)} \left( C_1 e^{-\sqrt{\delta^2 - \omega_0^2}(t-t_0)} + C_2 e^{\sqrt{\delta^2 - \omega_0^2}(t-t_0)} + c.c. \right)$ where $C_1$ and $C_2$ are constants, with

$\omega_0 = \sqrt{\kappa(B_{em}(\overline{n}+1) + B_{abs}\overline{n})}$ and $\delta = \frac{1}{2}(B_{em}X / \overline{n} + (B_{em} + B_{abs})\overline{n})$. For the case of $\overline{n} >> 1$ and

a large negative dye-cavity detuning $\Delta = \omega_c - \omega_{ZPL} < -k_B T/\hbar$ relevant here, we have $B_{em} >> B_{abs}$

and $\overline{M}_e >> \overline{n}$, so that the formulas reduce to the simplified forms given in the main text. In our experiment, we monitor the grand canonical system undergoing thermally (and quantum-mechanically) driven statistical fluctuations under stationary conditions by analyzing the second-order correlation function $g^{(2)}(\tau) = \frac{\overline{n(t) \cdot n(t+\tau)}}{\overline{n(t)} \cdot \overline{n(t+\tau)}}$ of condensate mode photons. The open grand canonical system is reminiscent of a stochastically driven damped harmonic oscillator as readily seen when writing Eq. (1) of the main text as a second order differential equation,

yielding the quoted eigenvalues $\lambda_{1,2}$. The expected second-order coherence function of

condensate photons can be written as

$$g^{(2)}(\tau) = 1 + e^{-\delta\tau} \left[ \left( Y + i\sqrt{\omega_0^2 - \delta^2} Z \right) e^{-\sqrt{\delta^2 - \omega_0^2}\tau} + \left( Y - i\sqrt{\omega_0^2 - \delta^2} Z \right) e^{\sqrt{\delta^2 - \omega_0^2}\tau} \right]. \qquad (S2)$$

Other than in the more compact formula given in the main text, the constants introduced in eq.

S2 (Y and Z) are real numbers on both sides of the phase transition, which is helpful for the fitting procedure.

Furthermore, from eqs. S1 we can readily determine the steady state values for the photon number $\overline{n}$ and the population of the upper electronic states $\overline{M}_e$. In the limit of small losses (

$\kappa << B_{abs}\overline{M}_g$) relevant here, i.e., photons are reabsorbed (and thermalized) faster than they

leave the cavity, eq. 1a in the steady state limit yields $\overline{M}_e \cong \overline{M}_g \cdot B_{abs} / B_{em}$. That is, for

$B_{em} >> B_{abs}$ we have $\overline{M}_e << \overline{M}_g$ so that $\overline{M}_g \cong M$. Using eq. S1b we furthermore find

$\overline{n} \cong R_p M / \kappa$, which expresses that in the open system the average photon number in the cavity

is tunable by the pumping strength. Typical experimental numbers for a cutoff wavelength

$\lambda_c \cong$ 571.3 nm, corresponding to a dye-cavity detuning $\Delta = \omega_c - \omega_{ZPL} \cong$ -4.1 $k_BT/\hbar$ for rhodamine dye, are $B_{em} \cong$ 23.9 kHz and $B_{abs} \cong$ 420 Hz, and a cavity loss $\kappa \cong$ 2.3 GHz. For a molecule number M$\cong 5 \cdot 10^9$, we have $\overline{M}_e \cong 8.8 \cdot 10^7$. For typical condensate photon numbers, which are up to $\cong 5 \cdot 10^5$, the limits given above are fulfilled with good accuracy.

The number statistics of the photon condensate is characterized by the crossing of two eigenvalues of the matrix A (eqs. (1-2) in the main text) characterizing the fluctuations in the system. Due to the driven-dissipative nature of the system, the matrix A=A(s) describing the two-component system of photons coupled to the dye reservoir is non-Hermitian. Here s denotes a set of system parameters; relevant to the described measurements are the average photon number $\overline{n}$, which experimentally is controlled by the pump power, the cutoff wavelength, and the dye concentration, and we have $\delta(s)$ and $\omega_0(s)$ damping rate and oscillation frequency respectively. The eigenvalues of the non-Hermitian 2x2 matrix A(s) can be either both real or both complex: Two real eigenvalues mean biexponential damping of the fluctuations, while complex eigenvalues mean damped oscillatory behavior. A crossing of the two eigenvalues occurs in the complex plane for a certain value of the parameters s=$s_{EP}$, see also the discussion in the main text. This is the exceptional point where the two eigenvalues coalesce. The gap between the eigenvalues $\lambda_1$ and $\lambda_2$ of the non-Hermitian matrix A(s) opens in the real or the imaginary plane on different sides of the exceptional point.

At the exceptional point, small changes of s can lead to qualitative changes of the state one is in, i.e. biexponential or oscillatory decaying respectively. Critical behavior of common (equilibrium) phase transitions is here rather replaced by the sensitivity of the phase one is in to changes in the control parameter s, which becomes large near the excepitional point at $\delta = \omega_0$, which we understand as the point of the phase transition. On the other hand, the condensate will remain in the biexponential or the oscillatory phase both for $\delta \gg \omega_0$ and for $\delta \ll \omega_0$ respectively upon small changes in s, i.e. when being deep in one of the corresponding phases.

As discussed in the main text, Fig. 2B represents the expected phase diagram in a three-dimensional plot, showing the biexponential and the oscillatory condensate phases and, for large losses (near $\kappa \cong B_{abs}M$, see Ref. 28), the crossover to lasing. One may write the condition $\omega_0 = \delta$ for the position of the exceptional point between the two dissipative phases in the form $\kappa / B_{em}\sqrt{\overline{M_e}} = \frac{1}{4}\left[\left(\frac{\sqrt{\overline{M_e}}}{\overline{n}}\right)^{3/2} + \left(\frac{\overline{n}}{\sqrt{\overline{M_e}}}\right)^{1/2}\right]$. When expressed in terms of the scaled parameters $\kappa' := \kappa / B_{em}\sqrt{\overline{M_e}}$ (scaled loss rate) and $\overline{\alpha} := \overline{n}/\sqrt{\overline{M_e}}$ (scaled photon number), this formula reduces to the universal form $\kappa' = \frac{1}{4}\left[\frac{1}{\alpha^{3/2}} + \alpha^{1/2}\right]$. The expected phase diagram thus collapses to the two-dimensional form used in Fig. 4A of the main text. In terms of the scaled variables, the theory values for the undamped oscillation frequency and damping constant can be expressed as $\omega_0 / B_{em}\sqrt{\overline{M_e}} \cong \cdot\sqrt{\kappa'\alpha}$ and $\delta / B_{em}\sqrt{\overline{M_e}} \cong \frac{1}{2}\left(\frac{1}{\alpha} + \alpha\right)$.

Data analysis

The experimental data for the second-order coherence function has been fitted with the theory prediction of eq. S2, which is valid on both sides of the phase transition, using $\omega_0$, $\delta$, $Y$, and $Z$ (all being real numbers) as fit parameters. Depending on the difference $\delta$-$\omega_0$ being positive or negative, the resulting second-order coherence function is biexponential or oscillatory, respectively. Experimental data for the difference $\delta$-$\omega_0$, along with corresponding uncertainties, is shown in Fig. 3A of the main text for different average photon numbers for one of the five recorded data sets, (see Table S1). The black lines in the top panels of Fig. 1C of the main text show the fit results for two individual correlation data measurements.

In the oscillatory phase ($\omega_0 > \delta$) the second-order coherence of eq. S2 takes the explicit form $g^{(2)}(\tau) = 1 + e^{-\tau/\tau_c}\left(C_1' \sin(\Omega_0 \tau) + C_2' \cos(\Omega_0 \tau)\right)$, with a decay time $\tau_c = 1/\delta$ and an oscillation frequency $\Omega_0 = \sqrt{\omega_0^2 - \delta^2}$, while in the biexponential phase ($\omega_0 < \delta$) it takes the form $g^{(2)}(\tau) = 1 + \left(C_1'' e^{-\tau/\tau_{c1}} + C_2'' e^{-\tau/\tau_{c2}}\right)$, with the two decay rates $1/\tau_{c1,2} = \delta \pm \sqrt{\delta^2 - \omega_0^2}$. Here, the constants $C_1', C_2', C_1''$ and $C_2''$ here again are real numbers.

Figures 3B,C of the main text show the variation of the observed decay times $\tau_{ci}$ and, for the case of the oscillatory phase, the oscillation frequency $\Omega_0$ on the average photon number (data set #1). For a fitting of such data, the expected variation of the undamped oscillation frequency $\omega_0 \cong \sqrt{\kappa B_{em} \bar{n}}$ and damping constant $\delta \cong \frac{1}{2}\left(B_{abs} M / \bar{n} + B_{em} \bar{n}\right)$ on the average photon number $\bar{n}$ was assumed, where the molecule number $M$ and the loss $\kappa$ are free fit parameters. For the form of the damping constant $\delta$ given here, we have used the relation $\bar{M}_e \cong M \cdot B_{abs} / B_{em}$ for the population of the upper electronic state of the dye molecules. The Einstein coefficients $B_{abs}$ and $B_{em}$ for the corresponding wavelength have been determined by extracting the coefficient $B_{abs}$ from absorption measurements in an independent spectroscopic measurement performed on a macroscopic dye cuvette (39), and determining $B_{em}$ for the corresponding wavelength using the Kennard-Stepanov relation $B_{em} = B_{abs}\ exp(-\hbar(\omega - \omega_{ZPL})/k_B T)$.

The data points in Fig. 4A of the main text summarize the results of correlation data recorded in all five measurement data sets, and Extended Data Table 1 gives the obtained fit results. In Fig. 4 of the main text, the experimental data is plotted for the scaled variables described in the previous section, allowing to combine results recorded at different cavity cutoff wavelengths in the single diagrams shown. Note that, given that both $\delta$ and $\omega_0$ are proportional to $B_{em}\sqrt{M_e}$, this scaling also holds for the oscillation frequency $\Omega_0$ and the inverse damping rates $1/\tau_{ci}$.

**Supplementary Text**

Microscopic derivation of the rate equations

The derivation of the rate equations S1 is detailed in Ref. 6. Briefly, the rate equations describing the dynamics of the photon condensate can be obtained via a Lindblad master equation that follows from a microscopic quantum-mechanical model (6,41,42) describing the dye molecules, their vibrational modes or phonons, and the coupling to the photon modes of the cavity. Due to fast collisions of the dye with solvent molecules, the phonon excitations may be considered to be in thermal equilibrium at ambient temperature and treated as a reservoir to be integrated out. The Hamiltonian (also known as the Tavis-Cummings-Holstein model) for a collection of $M$ molecules and multiple cavity modes reads

$$H = \sum_k \left( \omega_k - \omega_{\mathrm{ZPL}} \right) a_k^\dagger a_k + \sum_{m=1}^M \left[ \Omega\, b_m^\dagger b_m + \Omega\sqrt{S}\, \sigma_m^z \left( b_m + b_m^\dagger \right) + g \sum_k \left( a_k \sigma_m^+ + a_k^\dagger \sigma_m^- \right) \right],$$

(S3)

where the cavity modes, $a_k$, $a_k{}^\dagger$, have dispersion $\omega_k$, and the zero-phonon line of the dye molecules is defined by $\omega_{ZPL}$. $\sigma_m{}^z$ and $\sigma_m{}^\pm$ are the Pauli z-matix and the raising/lowering operators in the two-dimensional space of electronic ground and excited states of dye molecule $m$, respectively. The vibrational frequency is $\Omega$, and $S$ describes the coupling of the phononic oscillator potential to the electronic transition. Since the phonon position operators are $\hat{x}_m \sim \left( b_m + b_m^\dagger \right)$, we understand from $\sigma_m{}^z = \sigma_m{}^+ \sigma_m{}^- - \sigma_m{}^- \sigma_m{}^+$ that the effect of this term is a displacement of the center of the phononic oscillator, with the sign depending on whether the molecule is in the electronic ground or excited state. Finally, a Jaynes-Cummings coupling between photons and dye is included via the small parameter $g$.

The molecular part of the Hamiltonian can be diagonalized by means of a polaron transformation, which leads to an effective, nonlinear coupling between photons and molecules, mediated by the phonon excitations of the dye. It is then possible to bring $H$ into a form suitable for a perturbative expansion in $g$ while treating the molecular vibrations as a reservoir. After expanding to second order in $g$, and discarding inter-molecular correlations as well as the coherent, first-order contribution *(6)*, by making the well-known Born-Markov and secular approximations, one arrives at the master equation (6,27,42).

$$\dot{\rho}(t) = -\mathrm{i}[H_S, \rho(t)] + \sum_k \frac{\kappa}{2} \mathcal{L}[a_k]\rho(t) + \sum_{m=1}^{M} \left[ \frac{R_p}{2} \mathcal{L}[\sigma_m^+] + \frac{\gamma}{2} \mathcal{L}[\sigma_m^-] \right] \rho(t)$$
$$+ \sum_k \sum_{m=1}^{M} \left[ \frac{\Gamma_k^-}{2} \mathcal{L}[a_k^\dagger \sigma_m^-] + \frac{\Gamma_k^+}{2} \mathcal{L}[a_k \sigma_m^+] \right] \rho(t),$$

(S4)

where we have also included the photon resonator loss $\kappa$, the external pump rate $R_p$, and the imperfect quantum efficiency $\gamma$ of the molecules. The frequency-dependent, incoherent coupling coefficients $\Gamma_k^\pm = 2ReK(\pm[\omega_k - \omega_{\mathrm{ZPL}}])$, which give rise to the photon absorption and emission processes, respectively, are defined as

$$K(\omega) = g^2 \int_0^\infty \mathrm{d}\tau \ \mathrm{e}^{-(R_p+\gamma)\tau/2} \, \mathrm{e}^{\mathrm{i}\omega\tau} \left( \exp\left\{ -4S \left[ (1 - \cos\Omega\tau)\coth\frac{\beta\Omega}{2} + \mathrm{i}\sin\Omega\tau \right] \right\} - \exp\left\{ -4S\coth\frac{\beta\Omega}{2} \right\} \right).$$

(S5)

Note that for the weak external pumping employed in our experiments, the influence of the rapid relaxation of the reservoir on the broadening of the molecular spectrum described by $K(\omega)$ is much larger than any additional broadening due to $R_p$ and $\gamma$. It is, therefore, valid to assume that $\Gamma_k^{\pm}$ remain constant over the experimentally explored range of the external pumping strength. Alternatively, a quantum Langevin equation can be derived at the level of the master equation (S4).

The rate equations (cf. Eqs. 1) can now be obtained from the master equation (S4) by calculating the equations of motion for the time-dependent expectation values of the photon number in cavity mode $m$, $\langle n_k \rangle$(t), and the number of excited molecules, $\langle M_e \rangle$ (t). In terms of operators these equations read,

$$\partial_t \langle a_k^\dagger a_k \rangle = -\kappa \langle a_k^\dagger a_k \rangle - \frac{1}{2} \sum_{m=1}^{M} \left[ \Gamma_k^+ \langle a_k^\dagger a_k \left(1 - \sigma_m^z\right)\rangle - \Gamma_k^- \langle a_k a_k^\dagger \left(1 + \sigma_m^z\right)\rangle \right],$$
(S6)

$$\partial_t \langle \sigma_m^z \rangle = R_p \left(1 - \langle \sigma_m^z \rangle\right) - \gamma \left(1 + \langle \sigma_m^z \rangle\right) + \sum_k \left[ \Gamma_k^+ \langle a_k^\dagger a_k \left(1 - \sigma_m^z\right)\rangle - \Gamma_k^- \langle a_k a_k^\dagger \left(1 + \sigma_m^z\right)\rangle \right],$$
(S7)

where the average is $\langle A \rangle (t) = \mathrm{Tr}\{A\rho(t)\}$. Assuming that the molecules are all identical, one can drop the molecular index and replace the sum over the $M$ molecules in Eq. (S6) by a factor of $M$. Summing over $m$ also in Eq. (S7), we arrive at

$$\partial_t \langle n_k \rangle = -\kappa \langle n_k \rangle - \Gamma_k^+ \langle n_k \left(M - M_e\right)\rangle + \Gamma_k^- \langle \left(n_k + 1\right) M_e \rangle,$$
(S8)

$$\partial_t \langle M_e \rangle = R_p \left(M - \langle M_e \rangle\right) - \gamma \langle M_e \rangle + \sum_k \left[ \Gamma_k^+ \langle n_k \left(M - M_e\right)\rangle - \Gamma_k^- \langle \left(n_k + 1\right) M_e \rangle \right].$$
(S9)

A further simplification arises from the fact that for large dye reservoir, the occupation of a given cavity mode $m$ and the total molecule excitation number are uncorrelated, i.e. their correlation factorizes,

$$\langle n_k M_e \rangle \approx \langle n_k \rangle \langle M_e \rangle. \tag{S10}$$

This closes the set of equations S8. For a single photon mode $\langle n \rangle = \langle a_0^\dagger a_0 \rangle$, and with the ground-mode emission and absorption coefficients

$$
\begin{aligned}
B_{\text{em}} &= \Gamma_0^- , \\
B_{\text{abs}} &= \Gamma_0^+ ,
\end{aligned} \tag{S11}
$$

this represents the rate equations 1 and completes the derivation of the rate equations from the microscopic quantum mechanical model. Two further remarks are at place here. First, observe that the total excitation number $\langle X \rangle := \langle n \rangle + \langle M_e \rangle$, conserved without external driving or loss, obeys the much simpler equation

$$
\begin{aligned}
\partial_t \langle X \rangle &= -\kappa \langle n \rangle + R_p \left( M - \langle M_e \rangle \right) - \gamma \langle M_e \rangle \\
&= -\left( \kappa - (R_p + \gamma) \right) \langle n \rangle - (R_p + \gamma) \langle X \rangle + R_p M,
\end{aligned} \tag{S12}
$$

where the emission and absorption terms present in Eqs. (S8) and (S9) have canceled and only drive and loss terms remain. The fact that $\partial_t \langle X \rangle \neq 0$ distinguishes the driven-dissipative photon-dye system from a perfectly closed system and is ultimately responsible for the oscillatory second-order coherence dynamics. Second, in the closed limit $\kappa, R_p, \gamma \to 0$, the rate equations predict a steady-state Bose-Einstein distribution in the long-time limit (43)

$$\frac{\langle n_k + 1 \rangle}{\langle n_k \rangle} = \frac{\Gamma_k^+ \langle M - M_e \rangle}{\Gamma_k^- \langle M_e \rangle} = e^{\beta(\omega_k - \omega_{\mathrm{ZPL}})} \frac{\sum_l \Gamma_l^- \langle n_l + 1 \rangle}{\sum_l \Gamma_l^+ \langle n_l \rangle} \ , \tag{S13}$$

where we have used Eq. (S.6) for the first, and Eq. (S.7) and the Kennard-Stepanov relation

$\Gamma_k^+/\Gamma_k^- = e^{\beta(\omega_k - \omega_{\mathrm{ZPL}})}$ for the second equality. The first equality is important for the description of the phase diagram, as we will see in the following note.

Fluctuations around the steady state

Here we address the derivation of the dynamics of the photon-number fluctuations $\Delta n$

and the excitation-number fluctuations $\Delta X$, as described by Eqs. (1) and (2) in the main text.

Defining the deviations from the steady state as $\Delta n(t) := \langle n \rangle(t) - \bar{n}$, $\Delta M_e(t) := \langle M_e \rangle(t) - \overline{M_e}$ and

$\Delta X := \langle X \rangle(t) - \bar{X}$, the photon-number fluctuations obey the regression law

$$\begin{aligned}
\partial_t \Delta n &= \left( -\kappa - B_{\mathrm{abs}}(M - \overline{M_e}) + B_{\mathrm{em}} \overline{M_e} \right) \Delta n + \left( B_{\mathrm{abs}} \bar{n} + B_{\mathrm{em}}(\bar{n} + 1) \right) \Delta M_e \\
&\approx - \left( B_{\mathrm{abs}}(M - \overline{M_e}) - B_{\mathrm{em}} \overline{M_e} + \omega_0^2/\kappa \right) \Delta n + \left( \omega_0^2/\kappa \right) \Delta X,
\end{aligned} \tag{S14}$$

where we have introduced $\omega_0^2/\kappa = B_{\mathrm{abs}} \bar{n} + B_{\mathrm{em}}(\bar{n} + 1) \approx B_{\mathrm{em}} \bar{n}$. The excitation-number fluctuations follow analogously,

$$\partial_t \Delta X = -(\kappa - (R_p + \gamma)) \Delta n - (R_p + \gamma) \Delta X \approx -\kappa \Delta n. \tag{S15}$$

Taking the equilibrium distribution from (S13), we can write

$$B_{\mathrm{abs}} \left( M - \overline{M_e} \right) = B_{\mathrm{em}} \overline{M_e} + B_{\mathrm{em}} \frac{\overline{M_e}}{\bar{n}} \ , \tag{S16}$$

from which we find

$$\partial_t \Delta n \approx -2\delta\Delta n + \left(\omega_0^2/\kappa\right)\Delta X, \tag{S17}$$

with $2\delta = B_{\text{em}}\overline{M_e}/\overline{n} + \omega_0^2/\kappa$. As stated in the main text, Eq. (2), we therefore have in matrix form,

$$\partial_t \begin{pmatrix} \Delta n \\ \Delta X \end{pmatrix} = \begin{pmatrix} -2\delta & \omega_0^2/\kappa \\ -\kappa & 0 \end{pmatrix} \begin{pmatrix} \Delta n \\ \Delta X \end{pmatrix}. \tag{S18}$$

The eigenvalues of this non-Hermitian matrix determine the behaviour of the second-order coherence function via the quantum regression theorem, which in this case states that the density-density correlations of spontaneous fluctuations obey the same regression law as the relaxation towards the steady state after a small perturbation, as given by Eq. (S18). For completeness, a detailed proof of this statement is given in the subsequent section.

<u>Second order coherence dynamics</u>

Here we give the formal proof that Eq. (S18) also applies to the dynamics of the second-order correlation function. The time-dependent photon density-density or second-order correlation function is defined as (43)

$$g^{(2)}(\tau) = \left.\frac{\langle n(t+\tau)n(t)\rangle}{\langle n(t)\rangle^2}\right|_{t\to\infty} = \frac{\text{Tr}\left[a^\dagger a\, e^{\hat{\mathcal{L}}\tau}\tilde{\rho}_\infty\right]}{\text{Tr}\left[a^\dagger a \rho_\infty\right]^2}, \tag{S19}$$

where $\hat{\mathcal{L}}$ is the total superoperator defined by the master equation (S4), the steady-state density matrix is defined as $\rho_\infty = \lim_{t\to\infty}\rho(t)$, and we introduced an effective density operator defined by $\tilde{\rho}_\infty := a^\dagger a\rho_\infty$. We note in passing that, for the normal-ordered second-order correlation function, one would need to set $\tilde{\rho}_\infty = a\rho_\infty a^\dagger$. Defining also the corresponding effective average $\langle\langle A \rangle\rangle(\tau) := \mathrm{Tr} A e^{\hat{\mathcal{L}}\tau}\tilde{\rho}_\infty$, the second-order coherence becomes

$$g^{(2)}(\tau) = \frac{\langle\langle n \rangle\rangle(\tau)}{\bar{n}^2}.$$ 

(S20)

The effective averages $\langle\langle n \rangle\rangle$ and $\langle\langle M_e \rangle\rangle$ obey almost the same equations as $\langle n \rangle$ and $\langle M_e \rangle$, the only difference arising from the fact that the trace of the effective density matrix does not vanish, $\mathrm{Tr}\,\tilde{\rho}_\infty = \bar{n}$, which gives $R_p M\,\mathrm{Tr}[\sigma^-\sigma^+\tilde{\rho}_\infty] = R_p\,(\bar{n}M - \langle\langle M_e \rangle\rangle)$. Thus, one finds equations of motion formally analogous to the rate equations,

$$\partial_\tau \langle\langle n \rangle\rangle = -\kappa\langle\langle n \rangle\rangle - B_{\mathrm{abs}}\langle\langle n(M - M_e) \rangle\rangle + B_{\mathrm{em}}\langle\langle (n+1)M_e \rangle\rangle,$$

(S21)

$$\partial_\tau \langle\langle M_e \rangle\rangle = R_p\,(\bar{n}M - \langle\langle M_e \rangle\rangle) - \gamma\langle\langle M_e \rangle\rangle + B_{\mathrm{abs}}\langle\langle n(M - M_e) \rangle\rangle - B_{\mathrm{em}}\langle\langle (n+1)M_e \rangle\rangle.$$

(S22)

It is important to note that the form of these equations is independent of the operator ordering in the definition of $g^{(2)}(\tau)$ in Eq. (S19). Their structure is determined by $\hat{\mathcal{L}}$ rather than the *per se* arbitrary effective density matrix. What does change are the initial conditions. This, however, is negligible for the relatively large photon numbers relevant here, with the difference being on the order of $1/\bar{n}$. The relation between the rate equations and (S21), (S22) may also be understood a the density-matrix ansatz. If $\rho_\infty$ is represented as a diagonal matrix with elements $P_{nM_e}^\infty$, then $\tilde{\rho}_\infty$ will also be diagonal with elements $\tilde{P}_{nM_e}^\infty(\tau=0) = nP_{nM_e}^\infty$. These provide the initial conditions for a system of equations formally identical to that for $P_{nM_e}$.

To truncate and close the hierarchy of expectation values in Eqs. (S.19) and (S.20), the higher-order moments $\langle\langle M_e \rangle\rangle$ need to be factorized, which can be done using a Gaussian identity for $P_{nM_e}$ applied to the steady-state expectation values, i.e.

$$\overline{n^2 M_e} = 2\overline{n}\,\overline{nM_e} + \overline{M_e}\ \overline{n}^2 - 2\overline{n}^2\overline{M_e},\tag{S23}$$

where the average is

$$\overline{A} = \sum_{n=0}^{\infty}\sum_{m=0}^{M} A P_{nM_e}^{\infty}.\tag{S24}$$

However, since

$$\overline{nA} = \sum_{n=0}^{\infty}\sum_{m=0}^{M} nA P_{nM_e}^{\infty} = \sum_{n=1}^{\infty}\sum_{m=0}^{M} A \tilde{P}_{nM_e}^{\infty} = \langle\langle A\rangle\rangle,\tag{S25}$$

Eq. (S23) also implies,

$$\langle\langle nM_e\rangle\rangle = \overline{n}\langle\langle M_e\rangle\rangle + \overline{n}\ \overline{nM_e} + \overline{M_e}\langle\langle n\rangle\rangle - 2\overline{n}^2\overline{M_e}.\tag{S26}$$

Introducing furthermore the vector

$$\boldsymbol{g}(\tau) = \begin{pmatrix} \Delta g_n^{(2)} \\ \Delta g_{n,M_e}^{(2)} \end{pmatrix} = \begin{pmatrix} \langle\langle n\rangle\rangle - \overline{n}^2 \\ \langle\langle M_e\rangle\rangle - \overline{n}\overline{M_e} \end{pmatrix},\tag{S27}$$

Eqs. (S21) and (S22) become

$$\begin{aligned}
\partial_\tau \Delta g_n^{(2)} = \overline{n}\Big\{ &-\kappa\overline{n} - B_{\mathrm{abs}}\overline{n(M-M_e)} + B_{\mathrm{em}}\overline{(n+1)M_e}\Big\} \\
&- \kappa\Delta g_n^{(2)} - B_{\mathrm{abs}}\left[(M-\overline{M_e})\Delta g_n^{(2)} - \overline{n}\Delta g_{n,M_e}^{(2)}\right] + B_{\mathrm{em}}\left[(\overline{n}+1)\Delta g_{n,M_e}^{(2)} + \overline{M_e}\Delta g_n^{(2)}\right],
\end{aligned}$$

$$\tag{S28a}$$

$$\partial_\tau \Delta g_{n,M_e}^{(2)} = \overline{n} \left\{ R_p(M - \overline{M_e}) - \gamma \overline{M_e} + B_{\text{abs}} \overline{n(M - M_e)} - B_{\text{em}} \overline{(n+1)M_e} \right\}$$
$$- (R_p + \gamma) \Delta g_{n,M_e}^{(2)} + B_{\text{abs}} \left[ (M - \overline{M_e}) \Delta g_n^{(2)} - \overline{n} \Delta g_{n,M_e}^{(2)} \right] - B_{\text{em}} \left[ (\overline{n}+1) \Delta g_{n,M_e}^{(2)} + \overline{M_e} \Delta g_n^{(2)} \right].$$

$$(S28b)$$

Using the steady-state solutions of the rate equations to eliminate the curly brackets, one

arrives at the linear system

$$\partial_\tau \boldsymbol{g}(\tau) = \begin{pmatrix} -\kappa - B_{\text{abs}}(M - \overline{M_e}) + B_{\text{em}}\overline{M_e} & \omega_0^2/\kappa \\ B_{\text{abs}}(M - \overline{M_e}) - B_{\text{em}}\overline{M_e} & -(R_p + \gamma) - \omega_0^2/\kappa \end{pmatrix} \boldsymbol{g}(\tau).$$

$$(S29)$$

Under the various simplifying assumptions made in the main text ($B_{\text{em}} \gg B_{\text{abs}}$, validity of Eq.

S13), this general set of equations is equivalent to Eq. (S18). In this way, we have formally rederived that, for the assumptions made, the steady-state density-density correlations indeed follow the same dynamics as small fluctuations around the steady state.

| Data set | $\lambda_c$(nm) | $B_{\text{abs}}$(s$^{-1}$) | $\kappa$ (ns$^{-1}$) | M |
|---|---|---|---|---|
| 1 | 571.3 | 420 | 2.2(2) | $4.76(3) \cdot 10^9$ |
| 2 | 570.4 | 490 | 1.7(1) | $6.53(1) \cdot 10^9$ |
| 3 | 570.4 | 490 | 2.6(1) | $4.08(14) \cdot 10^9$ |
| 4 | 570.4 | 490 | 2.5(1) | $2.04(3) \cdot 10^9$ |
| 5 | 575 | 219 | 2.7(3) | $1.83(2) \cdot 10^9$ |

**Table S1:** Summary of experimental parameters and fit results for the five different data sets, recorded with different cavity cutoff wavelengths $\lambda_c$ (see second column) and dye concentrations. Both cavity loss $\kappa$ and dye molecular number M are fit results (final two columns). The Einstein coefficient for emission $B_{\text{em}}$ can be determined from the quoted Einstein coefficient for absorption $B_{\text{abs}}$ for the corresponding wavelength using the Kennard–Stepanov relation.